\documentclass[twocolumn,showpacs,amsfonts,amssymb,nobibnotes,aps]{revtex4-1}
\usepackage{amsmath}
\usepackage{latexsym}
\usepackage{amssymb}
\usepackage{graphicx}
\usepackage{textcomp}
\usepackage{hyperref}

\def\ba{\begin{eqnarray}}
\def\ea{\end{eqnarray}}
\def\be{\begin{equation}}
\def\ee{\end{equation}}

\def\bm{\begin{math}}
\def\me{\end{math}}

\newcommand{\dummy}

\begin{document}

\title {Temperature and Composition Dependence of Kinetics of Phase Separation in Solid Binary Mixtures}
\author{Suman Majumder and Subir K. Das$^{*}$}
\affiliation{Theoretical Sciences Unit, Jawaharlal Nehru Centre 
for Advanced Scientific Research, Jakkur P.O, Bangalore 560064, India}

\date{\today}
\begin{abstract}
We present results for the kinetics of phase separation in solid binary mixtures ($A_1+A_2$) from Monte Carlo simulations 
of the Ising model in two dimensions. The simulation results are understood via appropriate application of 
the finite-size scaling theory. At moderately high temperatures, for symmetric $(50:50)$ compositions of $A_1$ and $A_2$ particles the average 
size of the domains exhibit power-law growth with the exponent having the Lifshitz-Slyozov value of 
$1/3$ from very early time. However, our analysis shows that for low enough temperatures, 
the growth exponent at early time is smaller than the Lifshitz-Slyozov value. For composition dependence, we find 
that at moderate temperature, even for extreme off-critical composition, curvature dependent correction to the growth law is weak  
which is counterintuitive in this case that gives rise to droplet morphology. 
This is however consistent with the recent understanding on curvature dependence of surface tension. 
Results from rather general studies on the finite-size effects with the variations of temperature and composition have also been presented.
\end{abstract}
\maketitle

\section{Introduction}\label{3_intro}
~Understanding of the thermodynamic and dynamic properties of multi-component mixtures in the miscible as well 
as immiscible phases are of crucial technological importance \cite{Herlach,Alloy,3_BRAY,3_HAASEN,Onuki,3_WADHAWAN}. 
Nevertheless, it appears that some basic understandings 
for mixtures involving even two components ($A_1+A_2$) are rather incomplete, particularly certain aspects in nonequilibrium dynamics. 
In this work we study the kinetics of phase separation in a solid binary mixture and address questions primarily related to coarsening 
dynamics involving nanoscopic length scales.
\par
~When a homogeneously mixed binary mixture is quenched inside the miscibility gap it becomes unstable to fluctuations 
and starts evolving towards its new equilibrium state where $A_1$ and $A_2$ phases coexist. During this evolution, $A_1$- and $A_2$-rich 
domains form and grow with time until the system entirely phase separates \cite{3_BRAY,3_HAASEN,Onuki,3_WADHAWAN}. 
Such domain coarsening is a scaling phenomena where patterns at two different times are self-similar. 
This is reflected in the properties of various morphology characterizing 
functions, e.g., the two point equal time $(t)$ order-parameter correlation function, $C(r,t)$, 
obeys the scaling relation
 
\begin{eqnarray}\label{scaledCr}
C(r,t) &\equiv& \tilde {C}(r/\ell(t)),
\end{eqnarray}
where $\tilde {C}(x)$ is a scaling function, independent of $\ell(t)$, the average domain 
size. The function $\tilde {C}(x)$ provides information on the type of pattern formation during the evolution and is expected 
to be different for different situations \cite{3_BRAY} dictated, e.g., by the presence or absence of order-parameter conservation. Typically 
$\ell(t)$ has power-law dependence on time as \cite{3_BRAY,3_HAASEN,Onuki,3_WADHAWAN}
\begin{eqnarray}\label{powerlaw}
 \ell(t)\approx At^{\alpha},
\end{eqnarray}
where $\alpha$ and $A$ are respectively the growth exponent and amplitude. The value of $\alpha$ 
depends on the order-parameter conservation, system and order-parameter 
dimensions and other factors. Here, of course, we are dealing with phase separation where the order-parameter is conserved 
and for this particular problem, is a scalar quantity.
\par
~ For conserved order parameter, $\alpha$ depends upon the transport mechanism which is different 
in solids and fluids \cite{3_BRAY,3_HAASEN,Onuki,3_WADHAWAN}. The growth 
amplitude, in each of these cases, depends on temperature, composition and material under consideration. 
In case of solid mixtures the growth occurs via diffusive mechanism. In that case one can relate 
the rate of increase of $\ell(t)$ with the chemical potential ($\mu$) gradient 
and write \cite{3_BRAY,3_Huse} 
\begin{eqnarray}\label{dldt}
 \frac{d\ell(t)}{dt} \sim \lvert  \overrightarrow{\nabla} 
\mu \lvert \sim \frac{\sigma}{\ell(t)^{2}},
\end{eqnarray}
where $\sigma$ is the $A_1-A_2$ interfacial tension. Solving Eq. (\ref{dldt}) one obtains 
$\alpha=1/3$. This is referred to as the Lifshitz-Sloyozov (LS) law \cite{3_LS}. The LS behavior 
is the only growth law expected for phase-separating solid mixtures when there is prominent bulk diffusion. 
As we will discuss later, depending upon the value of $\ell(t)$ one may expect some correction to the 
LS law, of course. Also, dominant interface diffusion at very low temperature can give rise to different 
value of the exponent. In fluids, hydrodynamics being important, the growth should be much faster 
\cite{3_BRAY,Onuki,3_WADHAWAN,3_Siggia,3_Furukawa,3_Suman,3_Suman_J}. 
In this case, in addition to the diffusive growth, 
there exist two more regimes, at late times \cite{3_Siggia,3_Furukawa,3_Suman,3_Suman_J,AHMAD1}: 
first is the viscous hydrodynamic regime where domains grow linearly with 
time and the next is inertial hydrodynamic regime where one expects $\alpha=2/3$. For the sake of 
brevity we do not provide further details on fluid phase separation here.
\par
~Kinetics of phase separation in solid binary mixtures has been studied by implementing 
Kawasaki exchange kinetics \cite{3_kawasaki}
in the Monte Carlo (MC) simulations \cite{3_MCbook,3_Frankel} of the Ising model
\begin{eqnarray}\label{2_Ising}
 H = -J\sum_{<ij>}S_i S_j;~S_{i}=\pm1,~J>0,
\end{eqnarray}
where one can identify the spin $S_{i}=+1(-1)$ at lattice site $i$ with an $A_1$-particle ($A_2$-particle). In Eq. (\ref{2_Ising}) 
$<ij>$ stands for summation over the nearest neighbors (nn). An alternative method of studying the kinetics of phase separation 
in solid mixtures is via the coarse-grained dynamical equations, viz., the Cahn-Hilliard (CH) equation \cite{3_CH}
\begin{eqnarray}\label{2_Cahn-Hilliard}
 \frac{d\psi(\vec{r},t)}{dt}=-\nabla^{2}[\psi(\vec{r},t)+
\nabla^{2}\psi(\vec{r},t)-\psi^{3}(\vec{r},t)],
\end{eqnarray}
where $\psi(\vec{r},t)$ is a coarse-grained space- and time-dependent order parameter.
The CH equation, when supplemented with thermal noise, should be equivalent to MC simulations
of Kawasaki Ising model. Kinetics of phase separation in such simple but useful models have 
been studied mostly after quenching the systems to moderately low temperature along the critical composition line 
\cite{3_Huse,Amar,Grest,3_Suman1,3_Suman2}. Beyond that, it is also of significant importance to understand the effects of 
temperature ($T$) as well as composition \cite{Heermann} on the growth law. 
In this paper we will deal with these issues using the Kawasaki Ising model.
\par
The LS law ($\alpha=1/3$) for diffusive growth is expected to be valid for the situation when 
one has dominant diffusion in the bulk ($D_b$). In that case, variation of temperature (quench depth) is expected 
to bring changes in the growth amplitude $A$ only. On other hand, at very low temperature, diffusion along 
interfaces ($D_s$) may play the dominant role \cite{3_Langer,3_Furukawa1,3_Lebowitz,3_Gemmert}. 
It is pointed out that with the lowering of temperature a change in the exponent 
\cite{3_Langer,3_Furukawa1,3_Lebowitz,3_Gemmert} from $\alpha=1/3$ to a smaller value $1/4$ should happen. 
This can be justified via the following arguments \cite{3_Gemmert}. In this case, it may be useful to write the CH equation 
with concentration dependent mobility. A simple form for such mobility \cite{3_Gemmert} is 
\begin{eqnarray}\label{mobility}
M(\psi)= D_s \left[ 1-\beta \left (\frac{\psi^2}{{\psi_0}^2}\right) \right],
\end{eqnarray}
 where 
\begin{eqnarray}\label{beta}
\beta =1-\left (\frac{D_b}{D_s} \right )
\end{eqnarray}
and $\psi_0$ is the equilibrium domain magnetization. 
At moderate temperatures when $D_b$ and $D_s$ are comparable, one has nonzero mobility in the bulk. On the other hand, for very low $T$, 
it is argued that $D_b \ll D_s$ and in the bulk $\psi=\psi_0$ everywhere. This gives rise to negligible bulk mobility. However, since 
$\psi=0$ at the interfaces, one has nonzero mobility of particles there. Such dominant interface mobility may be responsible for 
a smaller exponent ($1/4$) at low temperature. From a dimensional analysis of the mobility dependent 
CH equation, Gemmert et al. \cite{3_Gemmert} worked out the crossover from $\alpha =1/3$ to $1/4$, as a function of $\beta$. 
Further, it is pointed out that this interface diffusion mechanism will provide $\alpha=1/4$ only at early time and 
in the large length scale limit the LS value will be recovered, even at low enough temperature.
\par
~In case of composition variation, as one moves away from the critical value, the 
domain pattern changes from interconnected structures to droplets of the minority species in the 
background of the majority sea. For fluid phase separation \cite{3_Stauffer,3_Siggia,3_Tanaka,3_Suman,3_Sutapa,3_Sutapa1}, 
this difference in morphology brings in striking and important change in the mechanism and thus the exponent of the growth law. 
In solid binary mixtures, however, one does not expect the mechanism to change due to the variation of composition. Here, an interesting 
and important objective could be to search for the correction to the scaling law \cite{3_Huse,3_Suman1,3_Suman2}, 
if any, due to the finite, say nanoscopic, radius of curvature of well defined droplets. Possibility of such correction at early time was discussed 
even for $50:50$ composition \cite{3_Huse}. However, recently \cite{3_Suman1,3_Suman2} it has been demonstrated 
that for symmetric composition, such finite time correction is negligible due to the flat boundary 
in an ``all time'' elongated interconnected domain structure. Away from critical composition, 
a possible reason for the curvature dependence of the growth exponent could be due to the 
correction in the surface tension in droplet geometry as \cite{Tolman,2_Wortis,Anisimov,2_Winter,2_Block,3_Das_Binder,3_Das_Binder1}
\begin{eqnarray}\label{sigma2}
 \sigma(\ell)=\frac{\sigma(\infty)} {1+2 \left( \frac{\delta}{\ell} \right) +2 \left (\frac{\ell_c}{\ell} \right)^2},
\end{eqnarray}
where $\delta$ and $\ell_c$ have dimensions of length. Another contribution can possibly 
come from the curvature dependence of the kinetic prefactor \cite{3_Huse} (not shown) in Eq. (\ref{dldt}). 
Here, we note that in Ising-like symmetric models, 
the Tolman length \cite{Tolman} ($\delta$) is absent \cite{2_Wortis,2_Block,3_Das_Binder,3_Das_Binder1}. 
\par
Athough the current interest in kinetics of phase separation lies in much more complicated systems  
\cite{2_Horbach,2_Horbach1,2_Horbach2,2_Mitchell,2_Bucior,2_Yelash,2_Yelash1,2_Das_Horbach,2_Das_Horbach1,2_Hore,2_Das_Stat}
with realistic interactions and boundary conditions, many of the basic facts,
 as discussed above, are still incompletely understood even for simple Ising model. Further, a major challenge 
in computer simulations of nonequilibrium systems is the finite-size effects \cite{Heermann,3_Suman1,3_Suman2}
that pose 
difficulty in extracting the growth exponent in the asymptotic limit. Thus understanding of finite-size 
effects is also of crucial importance.
\par
~In this work, we have undertaken a detailed study of kinetics of phase separation in solid binary mixtures 
via Monte Carlo (MC) simulations of Kawasaki exchange Ising model in two spatial dimensions ($d=2$). 
Our studies have particular focus in the small domain size limit. We present results on the effects of variation of temperature as 
well as that of composition both on the morphology and growth dynamics. In addition, we provide important 
quantitative information on the finite-size effects. To obtain information on the growth law, 
the corresponding correction at early time and the finite-size effects, we relied on finite-size scaling analysis 
\cite{3_MCbook,3_Fisher,3_Privman}, among other techniques.  
\par
The rest of the paper is organized as follows. In Sec.\ref{3_methods}, we describe the details of simulation 
and finite-size scaling method. In Section \ref{3_Results}, we present the results. Finally, we conclude the paper with a brief 
summary in Section \ref{3_Sec4}.
\section{Methods}\label{3_methods}
\subsection*{A. Method of Simulation }\label{2_detail}
~As already stated, we performed MC simulations \cite{3_MCbook,3_Frankel} of Ising model in $d=2$, keeping the 
order parameter conserved. This conservation in the dynamics was implemented  via the Kawasaki exchange mechanism 
\cite{3_MCbook,3_kawasaki} where a trial move consisted of interchanging positions between a randomly chosen pair of nn spins. 
Standard Metropolis algorithm \cite{3_MCbook} 
was used to accept or reject a move. Such exchange trials for $L^{2}$ pairs ($L$ being the linear dimension of a square 
lattice, in units of lattice constant $a$, used for the simulations) of spins form one Monte Carlo step (MCS). In all our simulations, 
periodic boundary conditions were applied in both $x$- and $y$- directions.  
\par
~By now it is clear that the central quantity of interest in this study is the average domain size $\ell(t)$. This can be calculated via the 
scaling properties of various morphological functions. 
In this work we have calculated $\ell(t)$ by exploiting three such functions \cite{3_BRAY}: (a) 
From the decay of the correlation function \cite{Plischke}
\begin{eqnarray}\label{cor}
C(r,t)=\langle S_{i}S_{j}\rangle -\langle S_i\rangle \langle S_j\rangle;~~r=|\vec{i}-\vec{j}|,
\end{eqnarray}
more precisely from the distance where $C(r,t)$ crosses zero for the first time 
(note that for such conserved dynamics $C(r,t)$ exhibits damped oscillation around zero before decaying to the later); 
(b) From the first moment of the structure factor $S(k,t)$ ($k$ being the 
wave vector) which is the Fourier transform of $C(r,t)$; (c) Finally, from the first moment, 
\begin{eqnarray}\label{moment}
 \ell(t)=\int d\ell_{d}~\ell_{d} P(\ell_{d},t),
\end{eqnarray} 
of the normalized domain-size distribution function $P(\ell_{d},t)$, where 
$\ell_d$ is the separation between two interfaces in $x-$ or $y-$ directions. Results from all three different methods 
are found to be consistent with each other for all temperatures and compositions, differing only by constant numerical factors. 
So, for the presentation purpose, unless otherwise mentioned, we will use results from only one of them, viz., method (c).
\par
~Note that the above mentioned methods of calculating $\ell(t)$ not always give the accurate measurment, 
particularly at high temperatures, due to the presence of noisy clusters of the size of equilibrium correlation length $\xi(T)$. 
Hence we have calculated $\ell(t)$ from noise free morphology by implementing a majority spin rule \cite{3_Suman1,3_Suman2,Das_multi}. In this method 
one replaces a spin at a particular lattice site with the sign of the majority of the spins involving the spins sitting at that site and its nn. 
Depending upon the temperature, one may need to improve the quality of noise filtering by including distant neighbors or increasing the 
number of such iterations.
\subsection*{B. Method of analysis}\label{2_fs_analysis}
~In this work, we have extensively used the finite-size scaling method \cite{3_MCbook,3_Fisher,3_Privman}
for the interpretation of simulation results. So, it will be useful to 
provide a discussion on this technique to improve readability of the paper. Following Ref. \cite{3_Suman2}, below we briefly outline the basic 
working concept. 
\par
~In critical phenomena \cite{Stanley}, the singularity of a quantity $Z$, in the thermodynamic limit, 
is quantified in terms of the reduced temperature 
$\epsilon=|T-T_c|/T_c$ ($T_c$ being the critical temperature) as 
\begin {eqnarray}\label{1_Z_div}
Z \approx Z_0 \epsilon^{z},
\end {eqnarray}
where $z$ is a critical exponent. For example, the correlation length $\xi$ diverges as 
\begin{eqnarray}\label{2_xi_eqn}
 \xi\approx \xi_0 \epsilon ^{-\nu}.
\end{eqnarray}
A central objective in critical phenomena is to obtain accurate 
information about the exponents. In finite systems, however, such divergences are restricted, e.g., $\xi$ cannot be larger than the system 
size $L$. In that case, to account for the size effect, one introduces a scaling 
function $Y$, to write \cite{3_Fisher,3_Privman}
\begin {eqnarray}\label{1_sc_ansatz}
 Z \approx Y(x)\epsilon^{-z},
\end {eqnarray}
where $x(=L/\xi)$ is a dimensionless scaling variable that provides information on the deficiency of system size with respect to 
the thermodynamic limit. Note that in the limit $x\rightarrow \infty$ ($L\rightarrow \infty$ or $\epsilon \gg 0$) there is no size 
effect and one should recover Eq. (\ref{1_Z_div}). Hence $Y$ should  approach $Z_0$ in that limit. 
Another important point is that $Y$ should be independent of system size (see the definition of $x$). So, as the working priciple, one acquires 
data for different values of $L$ and uses the exponents involved as adjustable parameters to obtain data collapse 
in $Y$ vs $x$ plot. The values that accomplishes this purpose describes the behaviors of the quantities correctly in the thermodynamic limit. 
In addition to adjusting the value of the critical exponents, sometimes \cite{Das_PRL} it may also be crucial to vary the background contribution 
(non critical part), coming from small length fluctuations, to obtain such data collapse. Here we note that the background 
parts usually have rather weak temperature dependence.
\begin{figure}[htb]
\centering
\includegraphics*[width=0.4\textwidth]{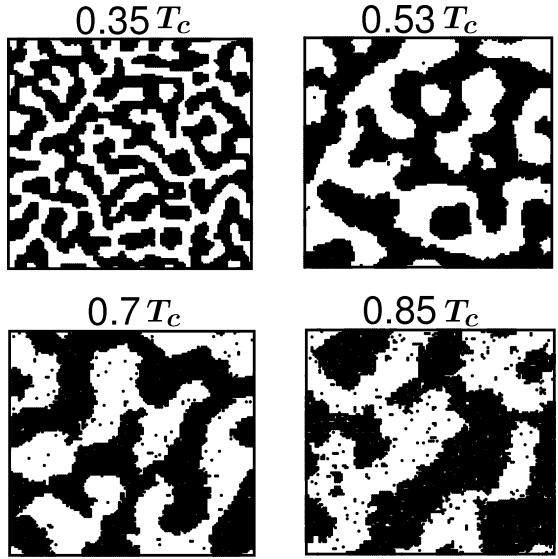}
\caption{\label{fig1} Snapshots obtained during the evolution of $2-d$ Ising model with $50:50$ composition and conserved 
dynamics, at $t=10^5$ MCS using systems of linear dimension $L=64$. Four different temperatures are included. Only $A_1$ particles are shown.}
\end{figure}
\par
~Similar scaling could be constructed in the nonequilibrium problems also, such as the present one, 
by identifying $\xi$ with $\ell$ and $\epsilon$ with $1/t$. We point out that the domain growth in an infinite system should be quantified 
as \cite{3_Suman1,3_Suman2}
\begin{eqnarray}\label{2_lteqn}
\ell(t')=\ell_{0}+At'^{\alpha},
\end{eqnarray}
where $\ell_{0}$ and $A$ are temperature dependent quantities. The length $\ell_0$ could be interpreted as the average 
cluster size when the system becomes unstable to fluctuations at time $t_0$ since the quench. Hence the 
measurement of time starts from there, i.e. $t'=t-t_0$. Note that the scaling part in Eq. (\ref{2_lteqn}) is only $At'^{\alpha}$, 
i.e, $\ell(t')-\ell_0$. This initial length $\ell_0$ is similar to the background in 
critical phenomena. Analogous to the weak temperature dependence of the backgrowund contribution in equilibrum critical 
phenomena, here we expect $\ell_0$ to be practically independent of time.
 Of course, if $\ell(t')$ is very large, 
subtraction of length $\ell_{0}$ does not
bring in major difference. However, in computer simulations, where one deals with finite systems,
if $\ell_{0}$ is not subtracted correctly one can end up with wrong conclusion. 
\par
Eq. (\ref{2_lteqn}) is true when there is no finite-size effects. For a finite system, 
analogous to (\ref{1_sc_ansatz}), one can write down the finite-size scaling form as 
\begin{eqnarray}\label{2_ltansatz}
 \ell(t')-\ell_{0}=Y(x)t'^{\alpha},
\end{eqnarray}
with
\begin{eqnarray}\label{x}
 x=\frac{\ell_{\max}-\ell_{0}}{t'^{\alpha}}
\end{eqnarray}
being the scaling variable. In Eq. (\ref{x}), $\ell_{\max}$ is the average length of domains in equilibrium that scales 
with $L$. Both in Eqs. (\ref{2_ltansatz}) and (\ref{x}), $\ell_{0}$ is 
subtracted to work only with the scaling parts. The limiting forms of $Y(x)$, which can be arrived at by using 
Eqs. (\ref{2_lteqn}), (\ref{2_ltansatz}) and (\ref{x}), are the following
\begin{eqnarray}\label{2_Yx}
 Y(x)&\approx& x,~ \mbox{for}~ x \rightarrow 0~(t'\rightarrow \infty, 
\ell_{\max} < \infty)
\end{eqnarray}
and
\begin{eqnarray}\label{2_Yx1}
Y(x)&=& A,~\mbox{for}~ x \rightarrow \infty~ (t' < \infty,~ \ell_{\max}\rightarrow\infty).
\end{eqnarray}
Learning the full form of $Y(x)$ would also be an interesting task, though challenging. Our objective in this analysis 
would be to interpret the simulation results from the behavior of the scaling function $Y$ which will be 
obtained from optimum data collapse by varying primarily $t_0$. Here we mention that we have varied $\alpha$ also to obtain optimum 
data collapse. In all the cases it appears that the value that serves the purpose best is $\alpha \simeq 1/3$.
\par
~In addition to the finite-size scaling, 
we have employed other methods of analysis as well which we will discuss subsequently while presenting the results.
\section{RESULTS}\label{3_Results}
~In this section we present results for the kinetics of phase separation in Ising model with critical composition for 
various depths of temperature quench.
\subsection*{A. Temperature Dependence}\label{3_Sec2}
\begin{figure}[htb]
\centering
\includegraphics*[width=0.4\textwidth]{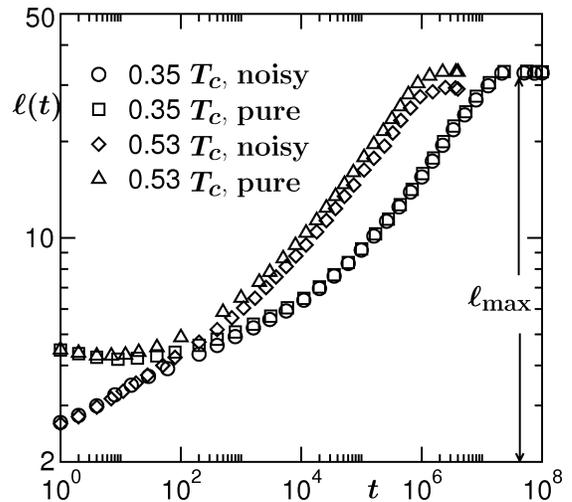}\\
\caption{\label{fig2} Log-log plots of average domain size, $\ell(t)$, as function of time, for $50:50$ composition. Results from 
two different temperatures are shown. For each of the temperatures we have included results from noisy and noise-free 
environments, as indicated. All results are obtained from the domain size distribution functions. 
The results were averaged over $100$ independent initial configurations with $L=64$. The equilibrium domain length 
$\ell_{\max}$ is defined in this picture.}
\end{figure}
In Fig. \ref{fig1}, we show the snapshots at $t=10^5$ MCS, from the evolutions 
at four different temperatures, after quenching a homogeneously mixed 
$50:50$ Ising system to the mentioned temperatures below $T_c$. Two observations are in 
order. First, the growth is faster for higher temperature. Second, the thermal noise, as expected, increases as one goes closer 
to $T_c$. As already mentioned, the later poses difficulty in accurate estimation of domain length $\ell$. 
To overcome this problem, as previously discussed, we calculated $\ell(t)$ from pure domain morphology 
after removing the noise. Quantitative justification of this is provided below. 
\begin{figure}[htb]
\centering
\includegraphics*[width=0.4\textwidth]{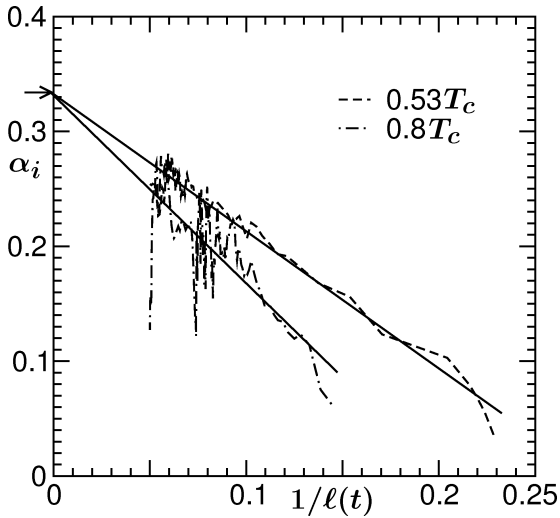}\\
\caption{\label{fig3} Plots of instantaneous exponents $\alpha_i$ vs $1/\ell(t)$, for two different temperatures, each with 
$50:50$ composition. The solid lines are guides to the eye. The arrow on the ordinate marks the value of $1/3$. 
All the results correspond to averaging over $100$ independent initial configurations and $L=64$.}
\end{figure}
\par
In Fig. \ref{fig2}, we show the plots of $\ell(t)$ vs $t$, on a double log scale, 
for two different temperatures. For both the temperatures we have included results from noisy as well as noise-free environments. 
For the lower temperature, of course, both the curves almost 
overlap with each other. This is due to very low level of noise. Notice that the equilibrium values, $\ell_{\max}$, of $\ell(t)$ 
(where the plots flatten) are different for the two temperatures 
for the calculation with noise. 
Here note that, since the results are presented for the same system size, $\ell_{\max}$ must be same for both the temperatures. 
Indeed, this is the case when the calculation is done from configurations without noise. This justifies the method, of course. 
However, at very early time, when the domains are very small, the noise removal exercise brings some undesirable feature. 
In this regime, along with noise, many domains also get removed. But it could be seen that 
during very early time, $\ell(t)$ for both the temperatures, calculated without removing the noise, match with each other, 
which is expected. In view of that, in our analysis, we will combine the 
$\ell(t)$ data with noise at early time (say, upto $t=10^2$ MCS for the temperatures in Fig. \ref{fig2})
and the noise-free data from the rest of the period. 
\par

\begin{figure}[htb]
\centering
\includegraphics*[width=0.4\textwidth]{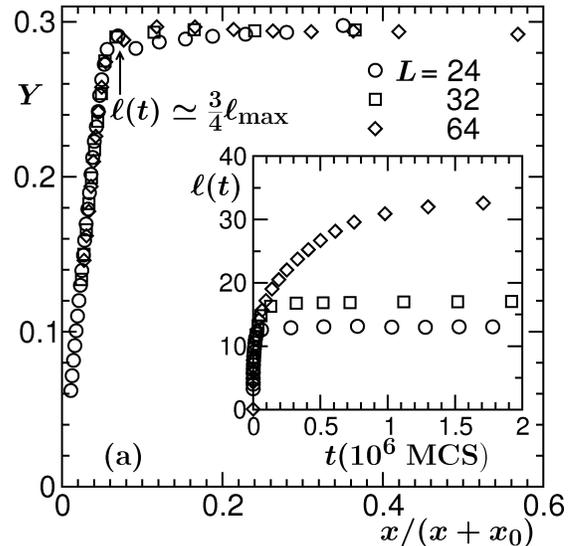}\\
\includegraphics*[width=0.4\textwidth]{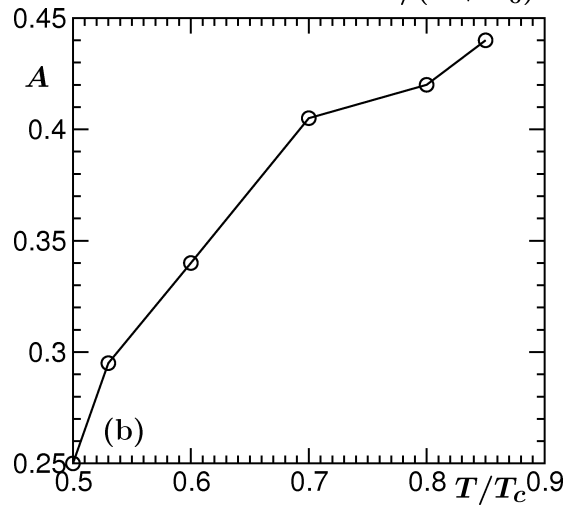}
\caption{\label{fig4} (a) Plot of the finite-size scaling function $Y$ vs 
$x/x+x_0$ with $x_0=5$, using three different system sizes, as indiacted. The results corresponds to $50:50$ composition at $T=0.53T_c$. 
The arrow there marks the onset of finite-size effect. Inset shows the plot of $\ell(t)$ vs $t$ for the data used in the main plot.
(b) Plot of growth amplitude $A$ vs temperature, for $50:50$ composition.}
\end{figure}
\par
~In addition to the methodological justification, as already mentioned, Fig. \ref{fig2} also provides 
the quantitative information about the fact that the equilibration 
occurs faster for higher temperature. This implies that, if the exponent is same, the amplitude of growth is larger for higher
temperature. We will investigate it in detail in the following.

\par
In Fig. \ref{fig3}, we have shown the instantaneous exponent $\alpha_i$, calculated via 
\begin{eqnarray}\label{alpha_i}
\alpha_i=\frac{d \ln \ell}{d \ln t},  
\end{eqnarray}
as a function of $1/\ell$, for two high temperatures. For the behavior of $\ell$ quoted in Eq. (\ref{2_lteqn}), one 
has \cite{Amar,3_Suman1,3_Suman2}
\begin{eqnarray}\label{alpha_i2}
\alpha_i=\alpha \left [ 1- \frac{\ell_0}{\ell} \right ].
\end{eqnarray}
The results for both the temperatures are consistent with this form showing convergence to $\alpha=1/3$, for 
$\ell \rightarrow \infty $. From the temperature dependent slopes of these numerical results we can make a guess about 
the background length $\ell_0$, and use it as a ``good'' trial value for the finite-size scaling analysis. 
From the exercise in Fig. \ref{fig3}, we obtain $\ell_0 \simeq 3.6$ and $4.5$ for $T=0.53T_c$ and $0.8T_c$, respectively. Here we 
note that the linear dependence of $\alpha_i$ on $1/\ell$ should not be mistaken with curvature dependent corection 
\cite{3_Suman1,3_Suman2}. This form merely appears due to nonzero value of $\ell_0$.
\begin{figure}[htb]
\centering
\includegraphics*[width=0.4\textwidth]{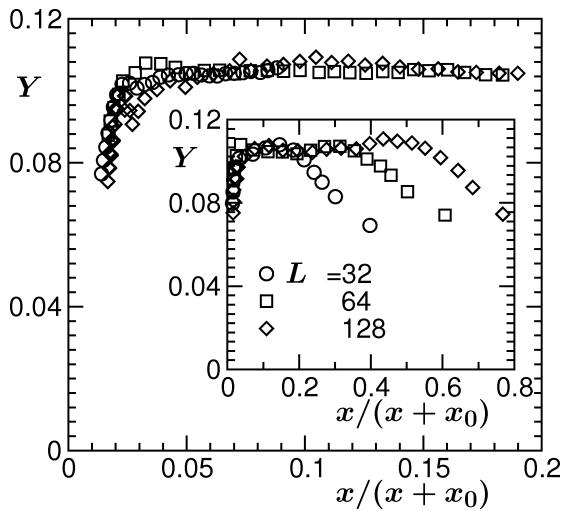}\\
\caption{\label{fig5} Finite-size scaling plot for $50:50$ Ising model using $\ell(t)$ data from three different system sizes 
at $T=0.35T_c$. The values of $\ell_0$ and $t_0$ are $3.9$ and $60$ MCS respectively. The value of $x_0$ used is $5$. 
The inset includes whole time range of simulations.}
\end{figure}
\par
~In Fig. \ref{fig4} (a) we show the exercise from the finite-size scaling analysis of data from different system 
sizes sizes at $T=053T_c$. (The original data on a direct plot are shown in the inset of this figure). 
This figure shows the plot of $Y$ as a function a function $x/(x+x_0)$ with $x_0=5$. 
Here $x_0$ was introduced to bring the infinity (corresponding to very early time) of the abscissa to a finite number. 
In this exercise we have fixed $\alpha$ to $1/3$ and varied $t_0$ to obtain optimum data collapse. The corresponding 
value of $\ell_0$ is in strong agreement with that obtained from Fig. \ref{fig3} using Eq. (\ref{alpha_i2}). 
Here, the flat part corresponds to the region unaffected by finite size of the systems. This flat behavior from very early time 
confirms about the absence of any significant length dependent correction to the LS value of the exponent. The deviation from this flat behavior 
provides us information about the onset of finite-size effects. At all temperatures we observe that 
this is rather weak. We will make quantitative statement about the size effect in the next subsection. 
Also note that the behavior of $Y$ at small $x$ is linear, consistent with Eq. (\ref{2_Yx}).
\par
In Fig. \ref{fig4} (b) we present a plot of amplitude $A$, extracted from the value of the ordinate 
for the flat parts of the finite-size scaling, vs $T$. One observes that $A$ 
monotonically increases with $T$. From the quality of data, however, it is difficult to figure out if there is 
any specific critical behavior of this quantity. 
\par
Uptil now, all the results correspond to reasonably 
high temperatures. Next we present results for low enough temperature to see if there is any signature of $\alpha=1/4$, 
due to dominant interface diffusion at early time.

\begin{figure}[htb]
\centering
\includegraphics*[width=0.4\textwidth]{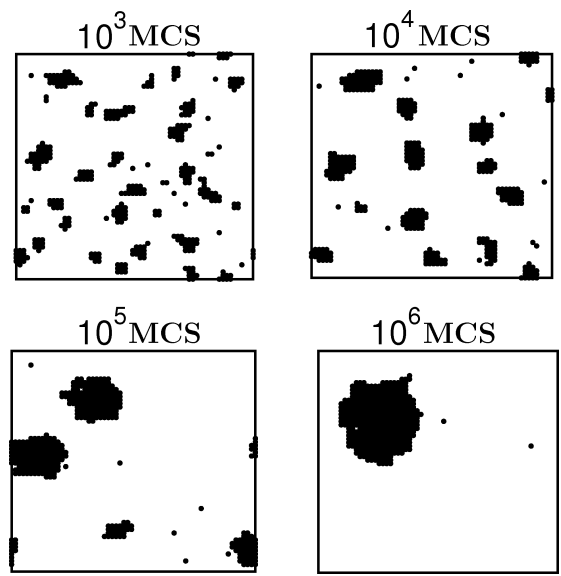}
\caption{\label{fig6} Snapshots obtained during the evolution of a $10:90$ Ising model at $T=0.53T_c$ 
in a square box of length $L=64$. Only $A_1$ particles are shown.}
\end{figure}
\par
From the nature of data at early time, for $T=0.35T_c$, in Fig. \ref{fig2}, it appears that the interface diffusion 
mechanism \cite{3_Langer,3_Furukawa1,3_Lebowitz,3_Gemmert} is perhaps giving rise to a smaller value of $\alpha$. 
In fact we have struggled to obtain even a reasonable finite-size collapse of data at early time, for this 
temperature, by fixing $\alpha=1/3$. This is demonstrated in Fig. \ref{fig5}. A good collapse of data is obtained only for $t>10^5$ MCS. 
The nonscaling behavior of the data in the range $t \epsilon [0,10^5]$ (see inset) and the trend 
of the early time data coming from the lower side indicate that the exponent is lower at the beginning. This could, 
possibly be identified with the ``expectation'' of $\alpha=1/4$ for lower temperature.
\par
~Further, our simulation results are also in nice agreement with the argument \cite{3_Gemmert} 
that at lower temperature there should be crossover from $\alpha=1/4$ to $1/3$. This, however, means that bulk mobility, 
which is negligible at early time becomes significant at late time. By examining Eqs. (\ref{mobility}) and (\ref{beta}), this 
is possible if the bulk order parameter decreases or $D_b$ increases with time. This is, we feel, counterintuitive and advocate for 
further studies to provide justification for that.
\begin{figure}[htb]
\centering
\includegraphics*[width=0.4\textwidth]{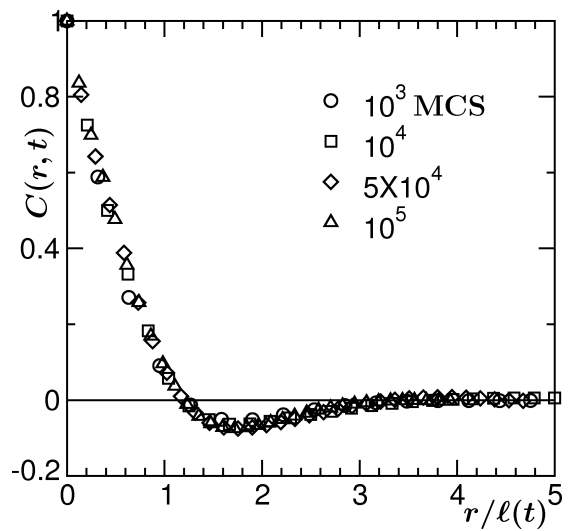}
\caption{\label{fig7} Scaling plots of the correlation functions for $10:90$ composition at $T=0.53T_c$. 
The results, for all the times, were obtained from a system of size $L=64$, after averaging over $100$ initial configurations.}
\end{figure}
\subsection*{B. Composition Dependence}\label{3_Sec3}
~In this subsection, we present results for morphology and growth from the MC simulations of Ising model with various different 
compositions of $A_1$ and $A_2$ particles. Here all the results are obtained at $T=0.53T_c$. 
\par
Fig. \ref{fig6} shows the snapshots during the evolution of a $10:90$ Ising system ($10$ $\%$ $A_1$ and $90$ $\%$ $A_2$) 
starting from a random initial configuration. Requirement of energy minimization and very off-critical composition 
restricts the domain geometry of the minority species to droplets. Since the hydrodynamics is unimportant in solid, 
growth of droplets, after their nucleation, takes place via simple concentration diffusion mechanism. As already stated, our objective here is to 
find out correction to the LS growth law, if any, at early time, due to finite radius of curvature of the droplets.

\par
In Fig. \ref{fig7}, we show scaling plots of the correlation function, for 
$10:90$ composition. Again, the length scale used here was obtained from the first moment of domain size distribution 
function. It is seen that one obtains good quality data collapse with the increase of time.
\par
\begin{figure}[htb]
\centering
\includegraphics*[width=0.4\textwidth]{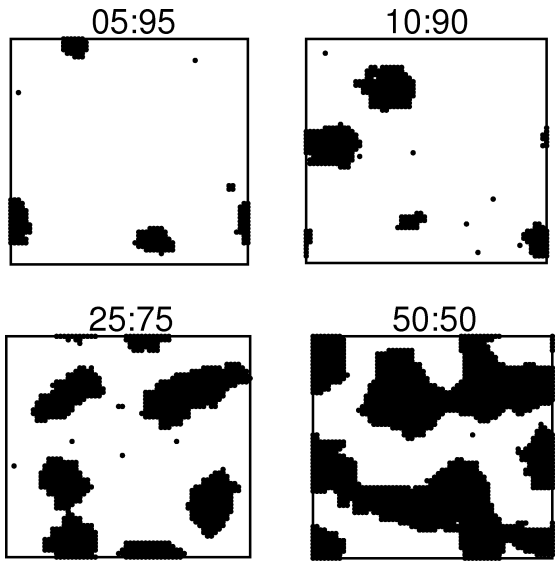}
\caption{\label{fig8} Snapshots from the evolution of Ising model at $T=0.53T_c$, 
for four different compositions at $t=10^5$ MCS, with $L=64$. Only $A_1$ particles are shown.}
\end{figure}

In Fig. \ref{fig8}, we show a comparative picture of snapshots from four different compositions at the same time ($t=10^5$ MCS). 
It is apparent that as one moves closer to the symmetric or critical composition, the growth is faster. This is is due to the fact 
that with the decrease of overall concentration of $A_1$ particles, they need to travel longer to get deposited on a domain.

\par
\begin{figure}[htb]
\centering
\includegraphics*[width=0.4\textwidth]{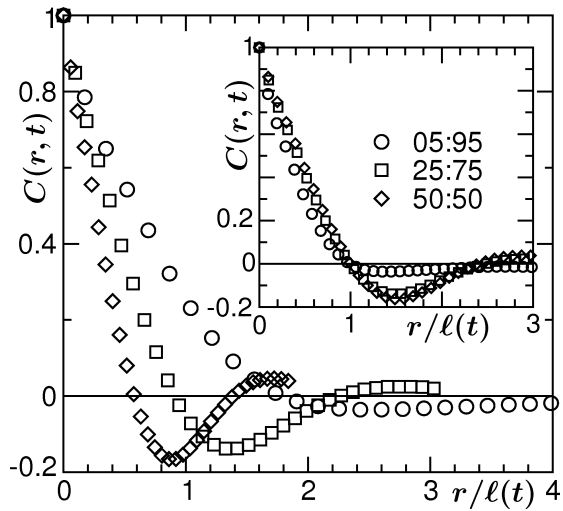}
\caption{\label{fig9} Scaling plots of $C(r,t)$ using data from three different compositions at $T=0.53T_c$, with $L=64$. 
Inset : Same as the main frame but here $\ell(t)$ was used from the first zero crossing of $C(r,t)$, as 
opposed to Eq. (\ref{moment}) in the main frame. All the results correspond to averaging over $100$ independent initial configurations.}
\end{figure}

In Fig. \ref{fig9}, we show the plots for correlation functions from four different compositions for the late time 
snapshots shown in Fig. \ref{fig8}. No scaling behavior is seen because of the method of calculation of $\ell(t)$. 
Note that we have used only the $A_1$-domains to calculate $\ell(t)$. But, from the definition of $C(r,t)$, it is clear that 
the information of both $A_1$-rich and $A_2$-rich regions are incorporated in it. Thus, we expect data collapse if in all the cases 
$\ell(t)$ is obtained from the decay of $C(r,t)$. This is demonstrated in the inset of this figure. However, 
there is mismatch in the amplitude of damped oscillation. 
This discrepancy is due to the fact that integration of $C(r,t)$ with respect to $r$ is proportional to the sum of the order parameter 
which is different for different compositions. Apart from that, reasonable collapse 
of data indicates that the basic structure is same in all the cases. 
\par
\begin{figure}[htb]
\centering
\includegraphics*[width=0.4\textwidth]{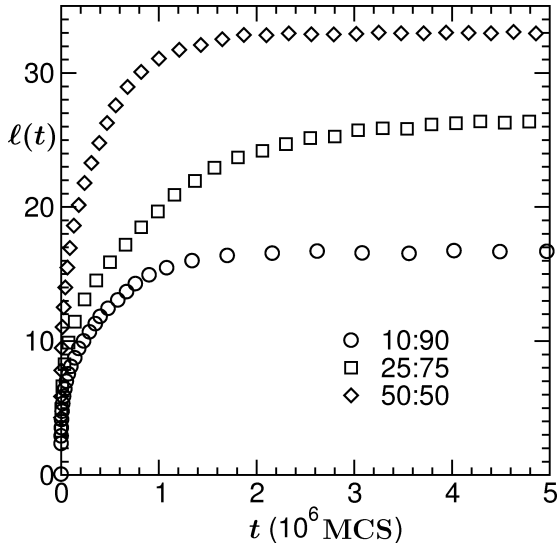}
\caption{\label{fig10} Plots of $\ell(t)$ vs $t$, at $T=0.53T_c$, for three different compositions. 
The system size for all the cases is $L=64$. Final results were obtained after averaging over 
$100$ independent initial configurations.}
\end{figure}
\begin{figure}[htb]
\centering
\includegraphics*[width=0.4\textwidth]{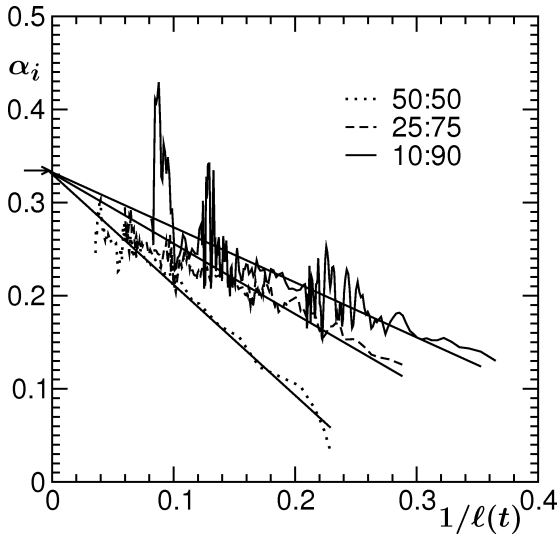}
\caption{\label{fig11} Plots of instantaneous exponent $\alpha_i$ vs $1/\ell(t)$ obtained from the plots in 
Fig. \ref{fig10}. The lines serve as guides to the eye. The arrow on the ordinate marks the LS value.}
\end{figure}

In Fig. \ref{fig10}, we show $\ell(t)$ as a function of time for three different compositions, as indicated. 
Interestingly it appears that the equilibration time $t_{eq}$ is a non-monotonic function of composition. 
Another observation is that the amplitude of growth decreases with increasing asymmetry of composition. 
This, combined with the fact that for extreme off-criticality smaller amount of $A_1$ particles need to 
assemble to reach equilibrium, can possibly explain the non-monotonic behavior of $t_{eq}$. 
In Fig. \ref{fig11}, we show the instantaneous exponents obtained for 
the plots in Fig. \ref{fig10}.
In all the cases it is seen that in the limit 
$\ell \rightarrow \infty$, there is a tendency of the data to converge to the LS value $\alpha=1/3$. 
\par
\begin{figure}[htb]
\centering
\includegraphics*[width=0.4\textwidth]{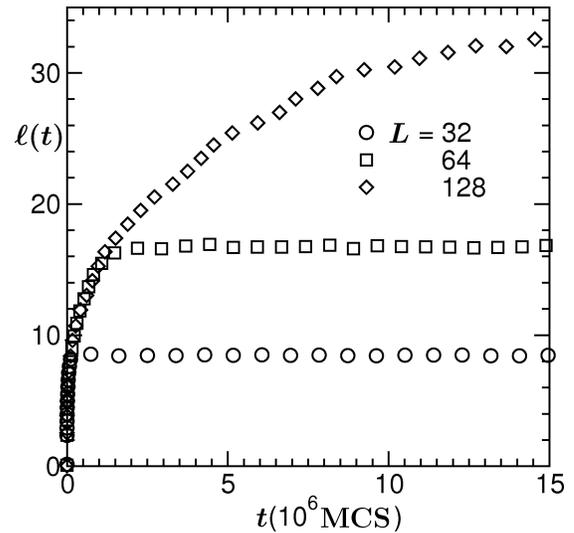}
\caption{\label{fig12} Average domain size, $\ell(t)$, is plotted vs $t$ for $10:90$ mixture. Results from three 
different system sizes at $T=0.53T_c$ are shown.}
\end{figure}
\begin{figure}[htb]
\centering
\includegraphics*[width=0.4\textwidth]{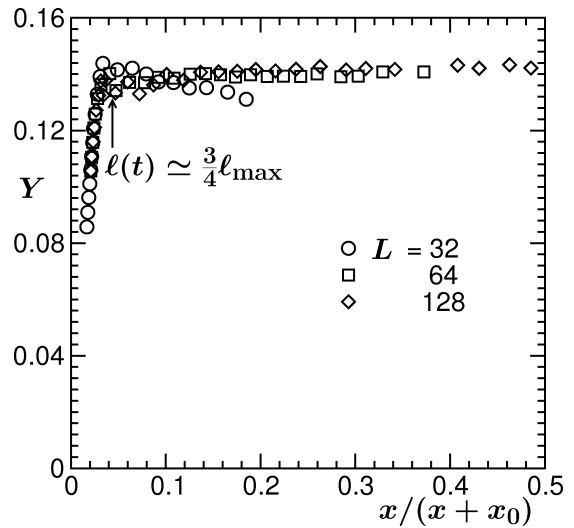}
\caption{\label{fig13} Finite-size scaling plot for data in Fig. \ref{fig12}. The values of $\ell_0$ and $t_0$ are respectively 
$1.8$ and $80$ MCS. Note that the value of $x_0$ used is $5$. The arrow marks the onset of finite size effect.}
\end{figure}

In Fig. \ref{fig12}, we show the plots of $\ell(t)$ vs $t$ from various different system sizes, for the composition 
$10:90$. Corresponding finite-size scaling analysis is demonstrated in Fig. \ref{fig13}. Best data collapse, that is presented 
here, was obtained for $\ell_0=1.8$ and $t_0=80$ MCS. This value of $\ell_0$ is consistent with the data in Fig. \ref{fig11}, 
for this composition.
Here again we have fixed $\alpha$ to the LS value $1/3$. Reasonable flat nature 
of the data from the begining confirms that the correction is negligible and if at all present, 
is buried in the statistical fluctuations of the simulation 
results. 
\begin{figure}[htb]
\centering
\includegraphics*[width=0.4\textwidth]{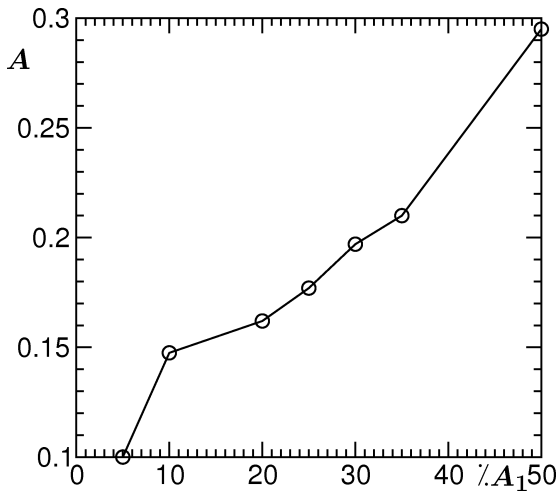}
\caption{\label{fig14} Plot of growth amplitude $A$ vs $\%$ of $A_1$ particles. All data points correspond to $T=0.53T_c$.}
\end{figure}
Again, from the ordinate of this flat region, we extract the amplitude of growth. This is plotted in Fig. \ref{fig14} 
as a function of composition.
\par 
Further, we have extracted information on the finite-size effects from the deviation of data from flat behavior in the 
finite-size scaling plots. In all the cases, both as function of temperature and composition, 
finite-size effects become visible at $\ell(t) \simeq \frac{3}{4} \ell_{\max}$. 
Here we note that, for asymmetric composition, if this is quantified as a fraction of the system size, this pre-factor 
will, of course, be much smaller than $\frac{3}{4}$. But we feel that it is more reasonable to quantify it with respect to the equilibrium 
length of the minority component, as it is done here. 
\par
Coming back to the point of curvature dependent correction, we provide a different analysis below. 
In our finite-size scaling analysis one may ask about the ambiguity in the estimation of $\ell_0$, 
however small it may be. To avoid such criticism, we search for a method where $\ell_0$ can be gotten rid of 
in a different mathematical way. Writing 
\begin{eqnarray}\label{3_ansatz}
 \ell(t)=\ell_0+At^{\alpha},
\end{eqnarray}
one obtains 
\begin{eqnarray}\label{3_dldt}
 \frac{d\ell(t)}{dt}=\ell'=A\alpha t^{\alpha-1}.
\end{eqnarray}
Since $t_0$ appeared microscopic in our previous analysis, here we have replaced $t'$ by $t$. 
By setting $\alpha=1/3$, one has
\begin{eqnarray}\label{3_lprime}
 \frac{1}{\ell'^{3/2}}=\left(\frac{3}{A}\right)^{3/2} t.
\end{eqnarray}
\begin{figure}[htb]
\centering
\includegraphics*[width=0.4\textwidth]{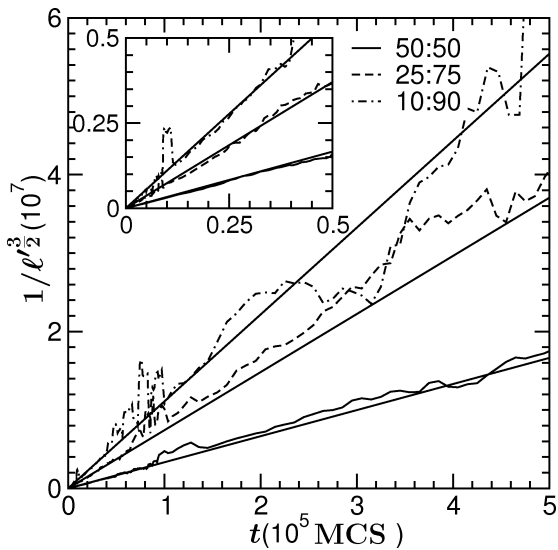}
\caption{\label{fig15} Plot of $1/\ell'^{3/2}$ vs $t$ for three different compositions at $T=0.53T_c$. 
The solid lines have slopes $\left(\frac{3}{A}\right)^{3/2}$. Inset shows a magnified picture involving only small $t$.}
\end{figure}
\par
In Fig. \ref{fig15}, we plot $1/\ell'^{3/2}$ vs $t$ for three different compositions, including the critical one. 
In all the cases, numerical results look consistent with linear behavior from very early time (see the inset), confirming 
about only weak correction to scaling. The solid lines there are guides to the eye for which we have used slopes 
$\left(\frac{3}{A}\right)^{3/2}$ by taking $A$ from Fig. \ref{fig14}. 
\par
Despite all these findings, we do not discard the fact that a correction is present. Our analysis might 
have failed to capture this aspect because of statistical fluctuation in the data combined with the fact that the corrections 
are of higher order in $1/\ell$ with very small value of prefactor. This argument stems from the curvature dependent interfacial tension 
for symmetric models, like the Ising one, with quadratic leading correction \cite{3_Das_Binder,3_Das_Binder1} [see Eq. (\ref{sigma2}) 
and subsequent discussion about nonexistence of $\delta$]. From 
equlibrium studies of a symmetric binary fluid that belongs to the Ising universality class of critical phenomena, 
it is observed that $\ell_c$ in Eq. (\ref{sigma2}) is very small at temperatures comparable to the present study. For temperatures 
significantly close to $T_c$, however, curvature dependent correction to the growth law may be significant. 
(We have seen such signature even for $50:50$ composition.) This statement can be 
justified from the finding that $\ell_c$ diverges as the equilibrium correlation length at criticality \cite{3_Das_Binder}. 
But, closer to $T_c$, computational study of kinetics of phase separation, particularly 
for off-critical composition, is limited by additional difficulties because of mixing of two 
diverging lengths, viz., the equilibrium correlation length $\xi$ and domain size $\ell$. So we leave it 
as a future task when we acquire better computational resources. 
\par
~We close this section with the following remarks. Irrespective 
of composition, at moderate temperatures, the correction to scaling appears very small and this contradicts and corrects our 
previous understandings about this phenomena in kinetics of phase separation. Further, the finite-size effects in all the cases 
appear rather weak.
\section{Conclusion}\label{3_Sec4}
~We have presented detailed results for the kinetics of phase separation in solid binary mixtures via 
Kawasaki exchange Monte Carlo simulation of Ising model in two dimensions. Results are understood via sophisticated 
finite-size scaling analysis and other methods. Important aspects related to the variation of quench depth 
and mixture composition are discussed. A particular focus was on the early time dynamics.
\par
~In case of critical composition, for temperatures above $0.5T_c$, it is observed that the growth law is consistent with 
Lifshitz-Slyozov (LS) law ($\alpha=1/3$) almost all along, with amplitude increasing as one moves closer to the critical point. 
For significantly low temperature, however, at early time, lower value of exponent 
was noticed which crosses over to the LS value only at much later time. In this case though 
the early time behavior is consistent with the prediction for interface diffusion mechanism, we feel that further studies are 
needed for appropriate understanding of growth over entire time range. 
\par
For the composition dependent case, our primary objective was to search for any significant correction to the scaling \cite{3_Huse}. 
The finding appears to be against it at least for the temperature considered. Our 
observation, however, is consistent with recent studies in the equilibrium context involving curvature dependent 
interfacial tension \cite{3_Das_Binder,3_Das_Binder1}. Nevertheless, in line of these equilibrium studies, we expect noticeable corrections for 
temperatures close to the critical value. Because of technical reasons, we, however, leave it as a future problem. 
\par
In addition to the growth law, we have presented results for the pattern 
formation also. These patterns were characterized via calculation of correlation functions. 
Lastly, our observation about the weak finite-size effects appears to be generic at all temperatures and compositions. 
\section*{Acknowledgment}\label{ack}
~The authors are grateful to the Department of Science and Technology, India, for financial support 
via grant number SR/S2/RJN-13/2009. SM also acknowledges Council of Scientific and Industrial Research, 
India, for financial support in the form of research fellowship. 
\par
${*}$ das@jncasr.ac.in

\end{document}